\documentclass[conference]{IEEEtran}
\IEEEoverridecommandlockouts

\usepackage{cite}
\usepackage{amsmath,amssymb,amsfonts}
\usepackage{algorithmic}
\usepackage{textcomp}
\usepackage{xcolor}
\usepackage{subfig}
\usepackage{soul}

\ifCLASSINFOpdf
  \usepackage[pdftex]{graphicx}
  \DeclareGraphicsExtensions{.pdf,.jpg,.png,.xps}
\else
  \usepackage[dvips]{graphicx}
  \DeclareGraphicsExtensions{.eps}
\fi

\def\BibTeX{{\rm B\kern-.05em{\sc i\kern-.025em b}\kern-.08em
    T\kern-.1667em\lower.7ex\hbox{E}\kern-.125emX}}
\begin{document} 
 
\title{Engineering and Experimentally Benchmarking a Container-based Edge Computing System \thanks{This work has been partially performed in the framework of mF2C project funded by the European Union's H2020 research and innovation programme under grant agreement 730929.}} 

\author{
\IEEEauthorblockN{Francisco Carpio, Marta Delgado and Admela Jukan}
\IEEEauthorblockA{Technische Universit{\"a}t Braunschweig, Germany, Email:\{f.carpio, m.delgado, a.jukan\}@tu-bs.de}
} 

\maketitle

\begin{abstract}
While edge computing is envisioned to superbly serve latency sensitive
applications, the implementation-based studies benchmarking its performance are
few and far between. To address this gap, we engineer a modular edge cloud
computing system architecture that is built on latest advances in
containerization techniques, including Kafka, for data streaming, Docker, as
application platform, and Firebase Cloud, as realtime database system. We
benchmark the performance of the system in terms of scalability, resource
utilization and latency by comparing three scenarios: cloud-only, edge-only and
combined edge-cloud. The measurements show that edge-only solution outperforms
other scenarios only when deployed with data located at one edge only, i.e.,
without edge computing wide data synchronization. In case of applications
requiring data synchronization through the cloud, edge-cloud scales around a
factor 10 times better than cloud-only, until certain number of concurrent users
in the system, and above this point, cloud-only scales better. In terms of
resource utilization, we observe that whereas the mean utilization increases
linearly with the number of user requests, the maximum values for the memory and
the network I/O heavily increase when with an increasing amount of data. 
\end{abstract}


\section{Introduction}

Edge computing systems are emerging today as the solution to bringing cloud
capabilities closer to the users thus not only reducing the latency as perceived
by users, but also the network traffic to remote cloud data centers. The latest
advances on lightweight virtualization (LV) techniques in form of containers and
serverless computing are especially gaining traction in edge computing. Unlike
the Kernel-Based Virtual Machines (KVM) commonly used in the cloud, containers
do not require manual server administration before launching the apps, and are
used as standalone and self-contained software packages that include all the
code and related dependencies necessary to run the app. In addition, container
platform solutions, such as Docker, are platform independent, just like KVMs,
allowing to execute apps independently from the operating system. As the name
implies, lightweight virtualization techniques are also able to run on
constrained devices, such as Raspberry Pis, making them a highly relevant IoT
solution.

When engineering an edge computing system, the issue about how to interconnect
distributed localities of computing nodes comes into play, and the performance
of the resulting system needs to be benchmarked in terms of latency, resource
utilization and scalability. While it is widely assumed that edge computing
outperforms the cloud in terms of latency, when engineering a real edge
computing system, this performance advantage is not given, or obvious. In fact,
many theoretical studies have pointed out that edge computing would show better
latency as long as sufficient inter-edge bandwidth and computing edge
capabilities are provisioned, which has not been tested real world systems yet.
In regard to the basic performance metrics, including latency, resource
utilization and scalability, there are still open questions regarding to the
real advantages that edge omputing solutions can bring.

We engineer to this end an edge and cloud system architecture with open source
software using Kafka at the edge for data streaming, Docker as application
platform and Firebase as realtime cloud database system. We then experimentally
benchmark the performance in terms of scalability, latency and resource
utilization, in scenarios where large number of amount of concurrent jobs in the
system are sending data concurrently. We compare three scenarios: cloud-only,
edge-only and combined edge-cloud. The results indicate that edge-only case
performs the best in terms of latency, as long as the application does not
require data synchronization between edge nodes. Otherwise, edge-cloud solution
is a better option until certain number of concurrent users in the system and a
certain amount of data; above that point, there is a clear disadvantage as
compared to cloud-only. The resource utilization measurements show that the mean
utilization is not affected by the amount of data and only the maximum values
for the memory and network I/O increase with the number of users when sending
larger amounts of data. 

The rest of the paper is organized as follows. Section II presents related work.
Section III describes the system architecture. The Section IV shows the
performance evaluation of the proposed system and Section V concludes the paper.

\section{Related Work}

\begin{figure*}[ht]
	\centering
	\vspace{-0.0cm}
	\includegraphics[width=0.95\textwidth]{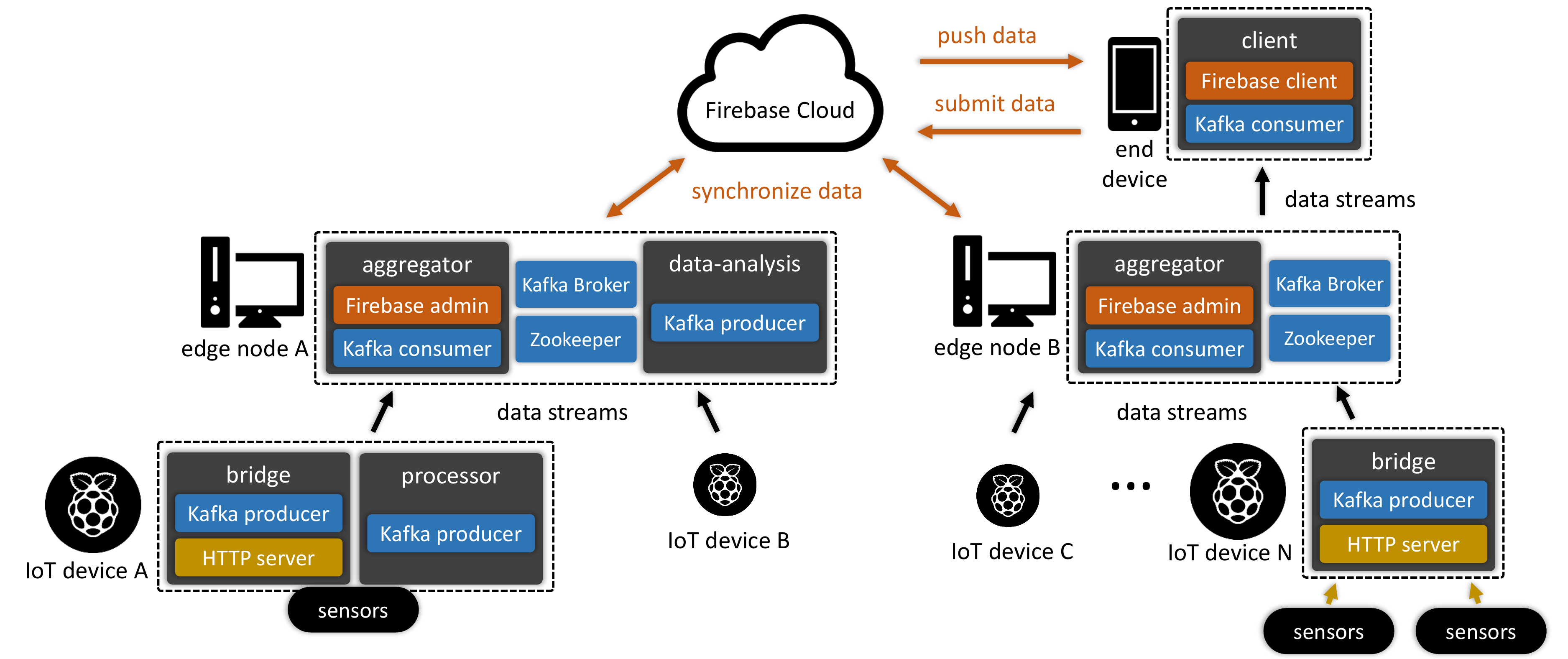}
	\caption{System Architecture}
	\label{sys_arch}
	\vspace{-0.1cm}
\end{figure*}

Edge computing is envisioned to computing, processing and storing data as close
as possible to the source of the data, which is critical to IoT systems
\cite{Pan2018a}, \cite{Zeydan2016}. Especially in the resource constrained edge
computing, traditional Kernel-Based Virtual Machines (KVM) have already been
replaced by lightweight virtualization solutions \cite{Ramalho2016}. Different
lightweight virtualization technologies, most notably containers or unikernels,
are used today as underlying tools to facilitating the fast adoption of edge
computing \cite{Morabito2018}. Despite significant engineering challenges,
including resource constraints, resilience and security \cite{Morabito2017,
Caprolu2019}, the LV solutions have been shown to exhibit good scalability,
resource and service management or fault tolerance, with a rather limited
overhead introduced \cite{Ismail2016, Bellavista2017}. Containerization is also
the most widely adopted technique and orchestration engine in today's cloud
systems, including Kubernetes or Docker Swarm being the de-facto standard of the
so-called \emph{serverless computing concept}, as adopted by AWS Lambda, Azure
Functions, Google Cloud Functions, etc, in form of various frameworks, such as
OpenFaas, Kubeless or OpenWhisk \cite{Palade2019}. Despite significant momentum,
the implementation-based studies benchmarking its performance are few and far
between. While lightweight virtualization on constrained devices have been shown
in a myriad of works as a promising solution for edge computing, once
engineered, the system needs to be benchmarked for the performance expected. For
the first time, we focus on that engineering and benchmarking process precisely,
and show how to experimentally evaluate the edge computing performance in three
real-world scenarios: cloud-only, edge-only and edge-cloud.

\section{System Architecture}

The reference architecture is shown in Fig. \ref{sys_arch} and follows the
modular microservice design principles. To this end, different architecture
components and modules can be combined to follow different configurations.
Starting from the bottom up, the architecture includes IoT devices
(single-boards, such as Raspberry Pis), over to the so-called edge nodes
(desktop computers, laptops, servers), the related end-devices (smartphones or
tablets) up to the traditional cloud service (in our implementation, Firebase).
IoT devices, having sensors attached directly to them, or receiving the data
from external sensors connected to it, act as data producers. In our system, we
have developed two docker ARM-based images: the bridge and the processor. The
bridge implements an HTTP server to receive data from other containers, for
instance, from the processor or from external sensors. By using a Kafka
producer, the bridge also sends data to Kafka Broker located, in this case, in
the edge node. The edge node runs the following containers: Kafka Broker,
Zookeeper, the Aggregator and Data-Analysis. Zookeeper and Kafka Broker work
together and are used as streaming messaging platforms. The Aggregator receives
data from Kafka by subscribing to it using a consumer, and stores the data into
Firebase through a Firebase Admin. Data-analysis as well as Processor are not
required for the main architecture to work but are the components designed to
performing some kind of processing on the data (for instance, machine learning)
to be later sent to the Aggregator. The cloud is implemented as a realtime
database instance using Firebase platform which flexibly stores data using JSON
tree structures.  Finally, the client is a Web-based module with both Firebase
and Kafka interfaces in order to receive data either using the cloud or the edge
nodes. While the detailed functionality and implementation of each module is out
of the scope of this paper, we here focus on the parts relevant to the
communication between the modules which will be later used to benchmark the
latency and other performance.

\subsection{Data Streaming}
In edge computing, efficient and reliable communication between different
modules is critical to achieve scalability. While different communication
protocols, following both client-server and publish-subscribe approaches, are
being used at the application layer, HTTP is being the most predominant one
\cite{Dizdarevic2019a} which is commonly used adopting the RESTful architectural
design as application programming interface (API). HTTP protocol, which was
designed as client-server based model for web browsers, is however not optimized
for managing large amounts of data stream messages due to excessive overhead.
Other publish-subscribe communication protocols, such as AMQP or MQTT can handle
scalability much better than HTTP and have been traditionally used in the cloud.
Despite this, we decided to use Apache Kafka, -- a known distributed streaming
platform based on the publish-subscribe model currently used by major industries
(including Netflix, Airbnb, Microsoft, Linkdin). With Kafka, in comparison with
traditional AMQP or MQTT, while the broker is not able to track the messages
received that have been acknowledged by the consumers, it is possible to achieve
high throughput by ingesting a large amount of data very quickly; for instance,
Kafka is proven to manage more than 20 million of events per second in Netflix's
system\footnote{S. Uttamchandani (2018, March 12). Lessons learnt from Netflix
Keystone Pipeline handling trillions of daily messages. Retrieved from
https://quickbooks-engineering.intuit.com/lessons-learnt-from-netflix-keystone-pipeline-with-trillions-of-daily-messages-64cc91b3c8ea}.
This tool together with processing tools like Apache Spark becomes a powerful
data processing solution.

\subsection{The Aggregator}

Since Kafka keeps track of all sent messages by persistent storage, and is also
built to run as a cluster, that can expand or contract, where data is replicated
internally to provide high availability and fault-tolerance, it is often
considered a database system. On the other hand, Kafka is usually not used as a
replacement of traditional database systems, but as support acting as commit
log. In this context, an interface is required between Kafka and the database
system, which we engineer herewith as Aggregator. The Aggregator is developed in
Java a consist of a Kafka consumer that is subscribed to all topics which data
need to be stored in the database. Since all data exchange in our system follows
the JSON specification where every message contains one JSON object, the
Aggregator only requires an identification field for every exchanged object.
These objects are then stored into the database through the Firebase Admin SDK
that the Aggregator also incorporates.

\subsection{Data Storage and Synchronization}

As previously mentioned, our system relies on a Firebase real-time database
system specifically designed to maintain data synchronization among multiple
clients. This database is NoSQL, and instead uses JSON trees structures to store
data, which provides more scalability and full flexibility when defining data
structures as compared to traditional SQL-based databases. The selection of
Firebase database system rather than other NoSQL available ones is basically
because of the ability of maintaining all clients synchronized automatically
removing the effort of developing periodic queries to the database. There is
also another feature that makes Firebase a suitable option for our architecture,
which is the ability of working in offline mode. Since the Aggregator uses the
Admin SDK, it first stores the data locally and later synchronizes in best
effort mode to the cloud and to other clients. These features provide high
flexibility to the Aggregator which is now able to work not only on devices with
reliable connectivity, but also on mobile devices with intermittent connectivity.

\begin{figure*}[ht]
	\centering
	\subfloat[Scenario 1: Cloud-Only]{\includegraphics[width=0.6\columnwidth]{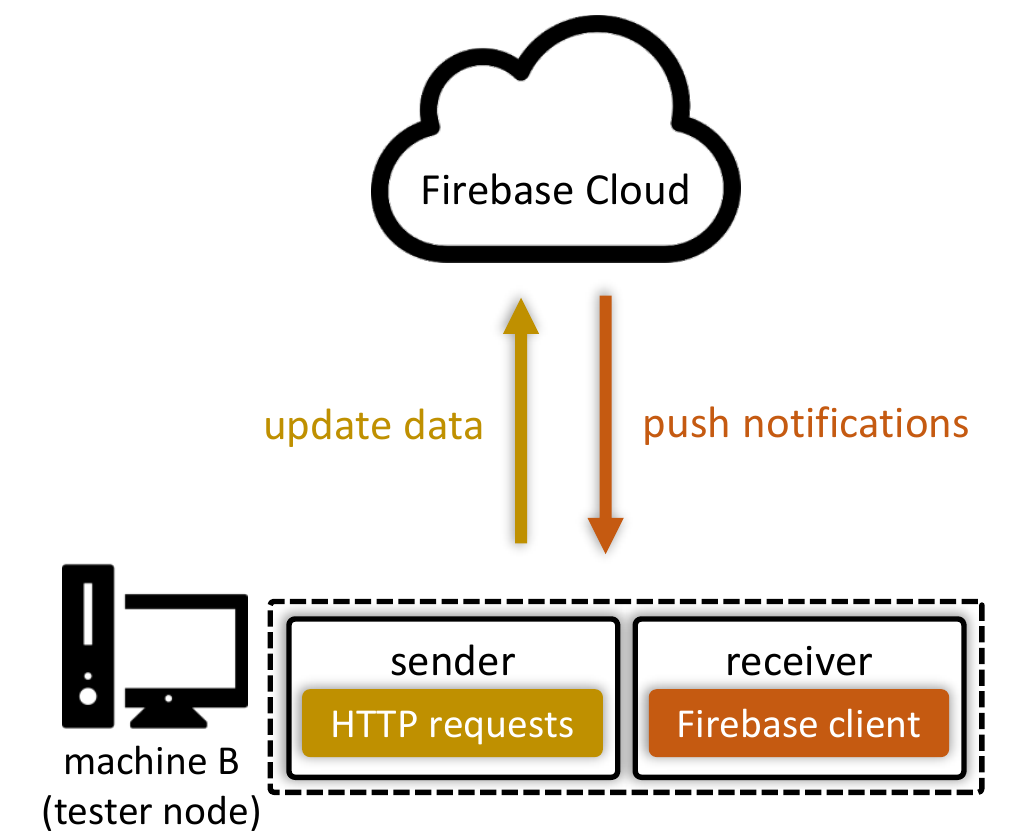}
	\label{scenario1}}
	\hfil
	\subfloat[Scenario 2: Edge-Only]{\includegraphics[width=0.6\columnwidth]{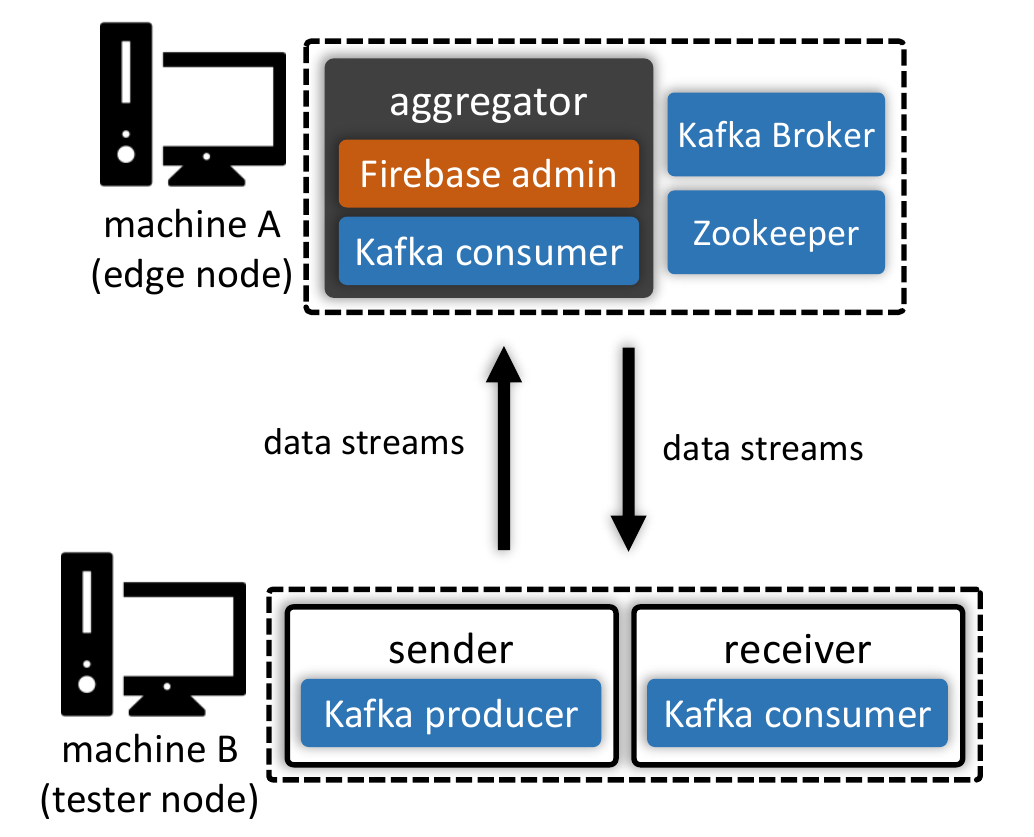}
	\label{scenario2}}
	\hfil
	\subfloat[Scenario 3: Edge-Cloud]{\includegraphics[width=0.8\columnwidth]{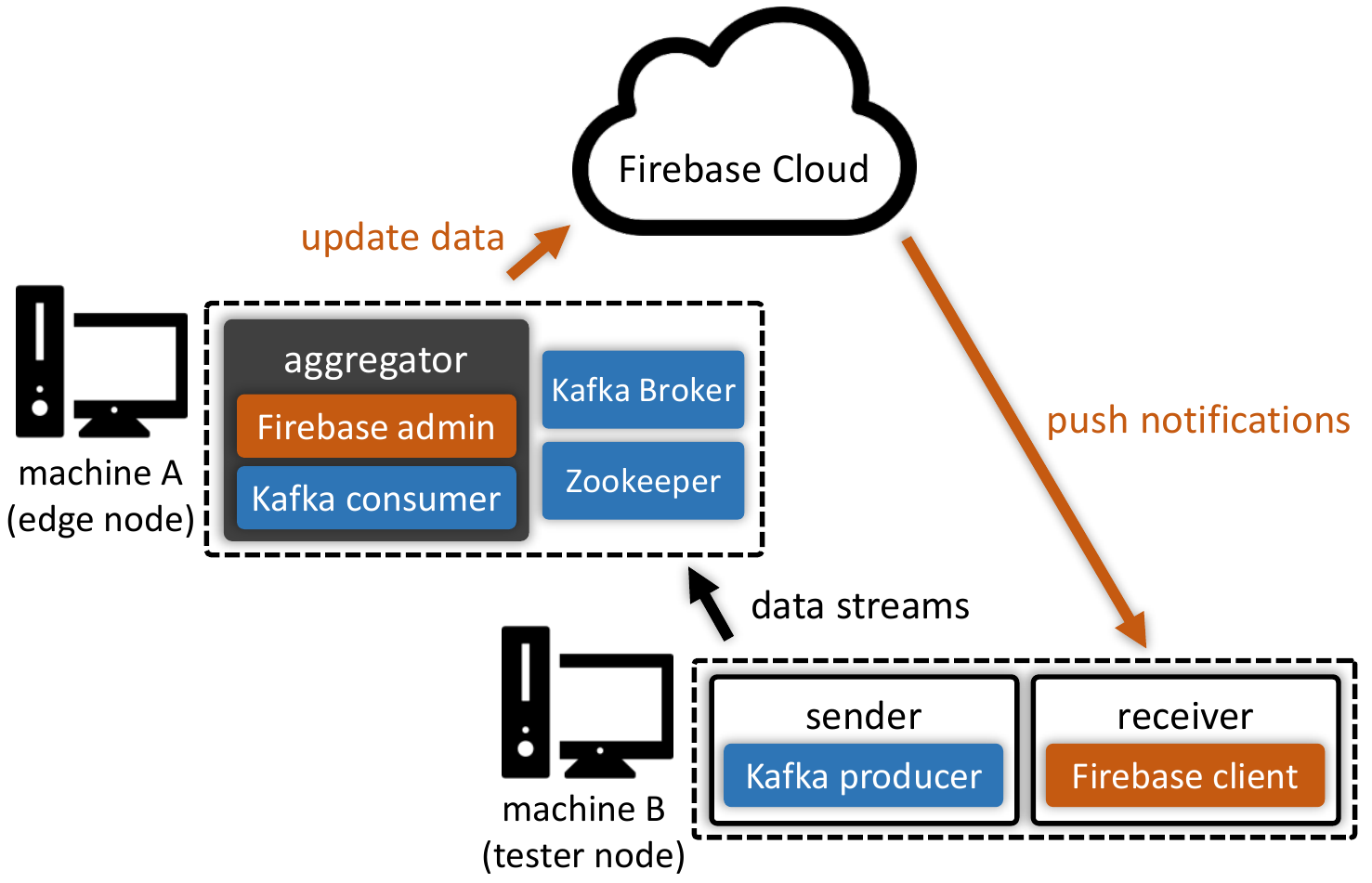}
	\label{scenario3}}
	\caption{Test scenarios}
	\label{scenarios} 
	\vspace{-0.1cm}
\end{figure*}

\section{Performance Evaluation}

Let us now introduce the testbed and the parameters used to perform the tests
and benchmark the performance. The testbed consists of two local desktop
computers (both with i5-7600 CPU, 16GB of RAM and SSD SanDisk X400 up to 540 and
520 MBps of reading and writing speeds, respectively), one desktop running as
edge node and the other one as tester node. Each computer has two Gigabit
Ethernet interfaces, where, in both cases, one is connected externally with 1
Gbps downlink and 500 Mbps uplink, and another one is used for the
interconnection between the two machines. The computer running as edge node is
running three Docker containers: the Aggregator, Kafka broker and Zookeeper. The
computer used for the tests runs a Java application with a Firebase client and
Kafka consumer to to receive data either from the cloud or from the Kafka Broker
depending on the scenario, and also stores the results of the measurements. In
the same machine, a Python script sends data with JSON format to either the edge
node using a Kafka producer or to the cloud using HTTP requests. In the Cloud,
we have a Firebase realtime database (located in Western Europe) that will store
the data sent by the Aggregator and will notify to the connected clients, in our
case the Java app running on the tester node. With these two machines and the
cloud, we evaluate three different scenarios shown in Fig. \ref{scenarios}. In
cloud-only scenario (see Fig. \ref{scenario1}), the tester node performs the
tests by sending HTTP requests to the cloud endpoint and uses the firebase
client to receive the updates asynchronously a soon as they are stored in the
database. In edge-only scenario (see Fig. \ref{scenario2}), the tester node
sends and receives data to and from the edge node using one Kafka producer and
one Kafka consumer. In a combined edge-cloud scenario (see Fig.
\ref{scenario3}), the tester node sends data to the edge node using the Kafka
producer and receive the updated data from the cloud using the Firebase client.

To benchmark the system, we performed tests to measure the total latency of the
system since the tester node sends data until the data is received in the same
machine. The scalability is measured in terms of how many concurrent users
(i.e., processes) the system support until the latency is too high that the
system becomes unusable. We also benchmark the resource utilization, including
CPU, memory and network utilization that are required by the containers running
on the edge node when performing the tests. For the different tests, the tester
node simulates different number of users (i.e., processes), from 100 to 1000, by
opening threats and sending different payload sizes depending on each case.
Every user individually sends 1000 requests with inter-arrival time following a
normal distribution with mean value 1 second. For each test, the database is
deleted and the containers at the edge node restarted in order to delete any
data from previous tests.

\subsection{Latency}

\begin{figure}[ht]
	\centering
	\subfloat[cloud-only - 1KB payload size]{\includegraphics[width=0.99\columnwidth]{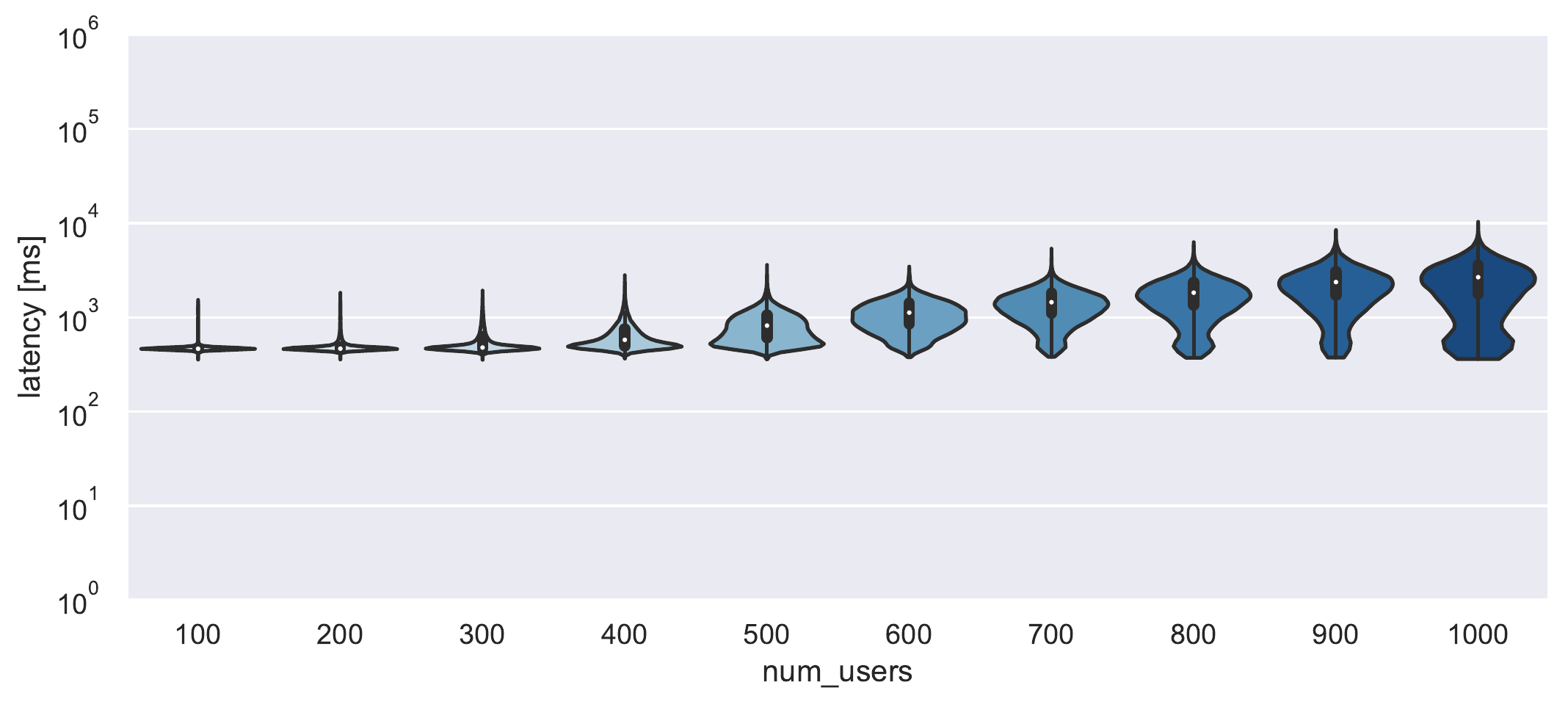}
	\label{fig_cloud_only_violinplot_1KB_latency}}
	\hfil
	\subfloat[edge-only - 1KB payload size]{\includegraphics[width=0.99\columnwidth]{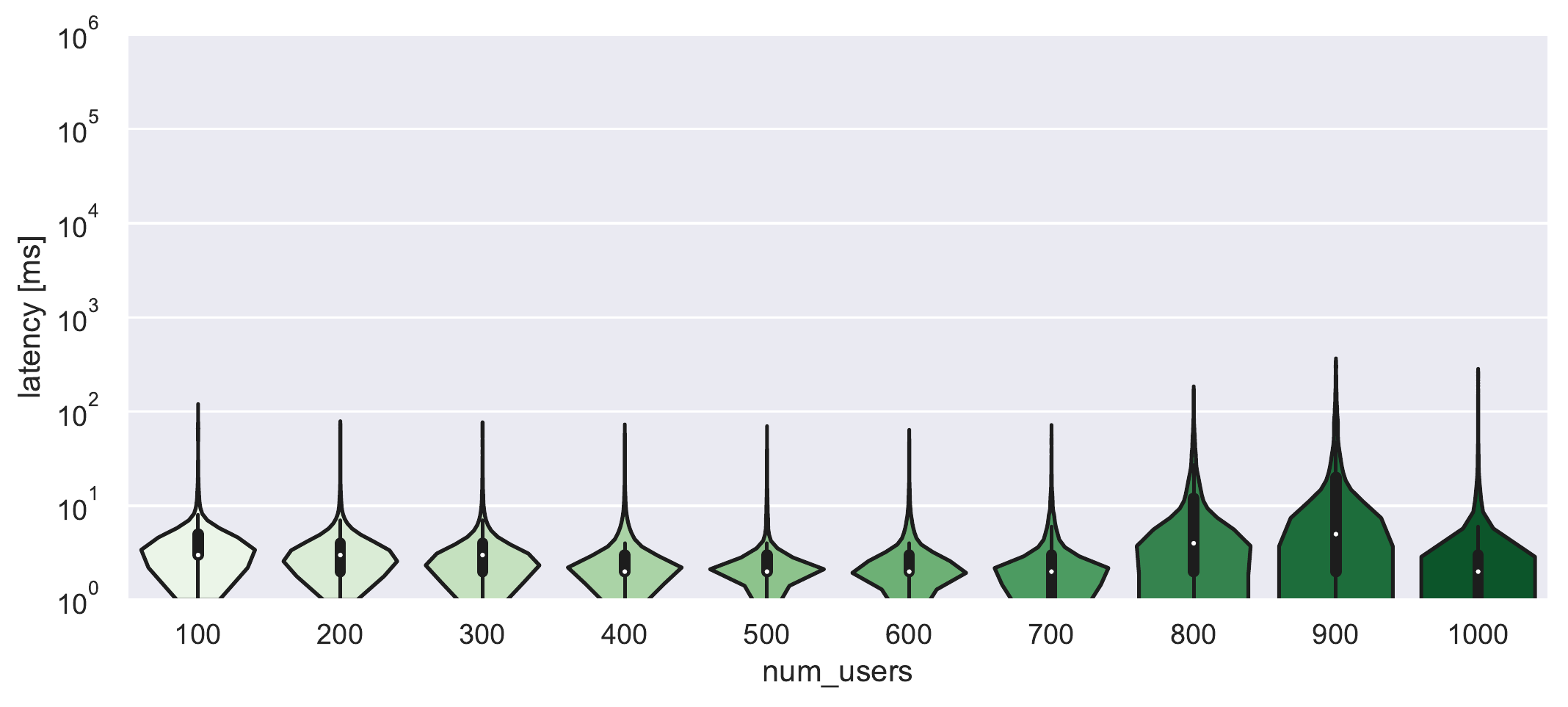}
	\label{fig_edge_only_violinplot_1KB_latency}}
	\hfil
	\subfloat[edge-cloud - 1KB payload size]{\includegraphics[width=0.99\columnwidth]{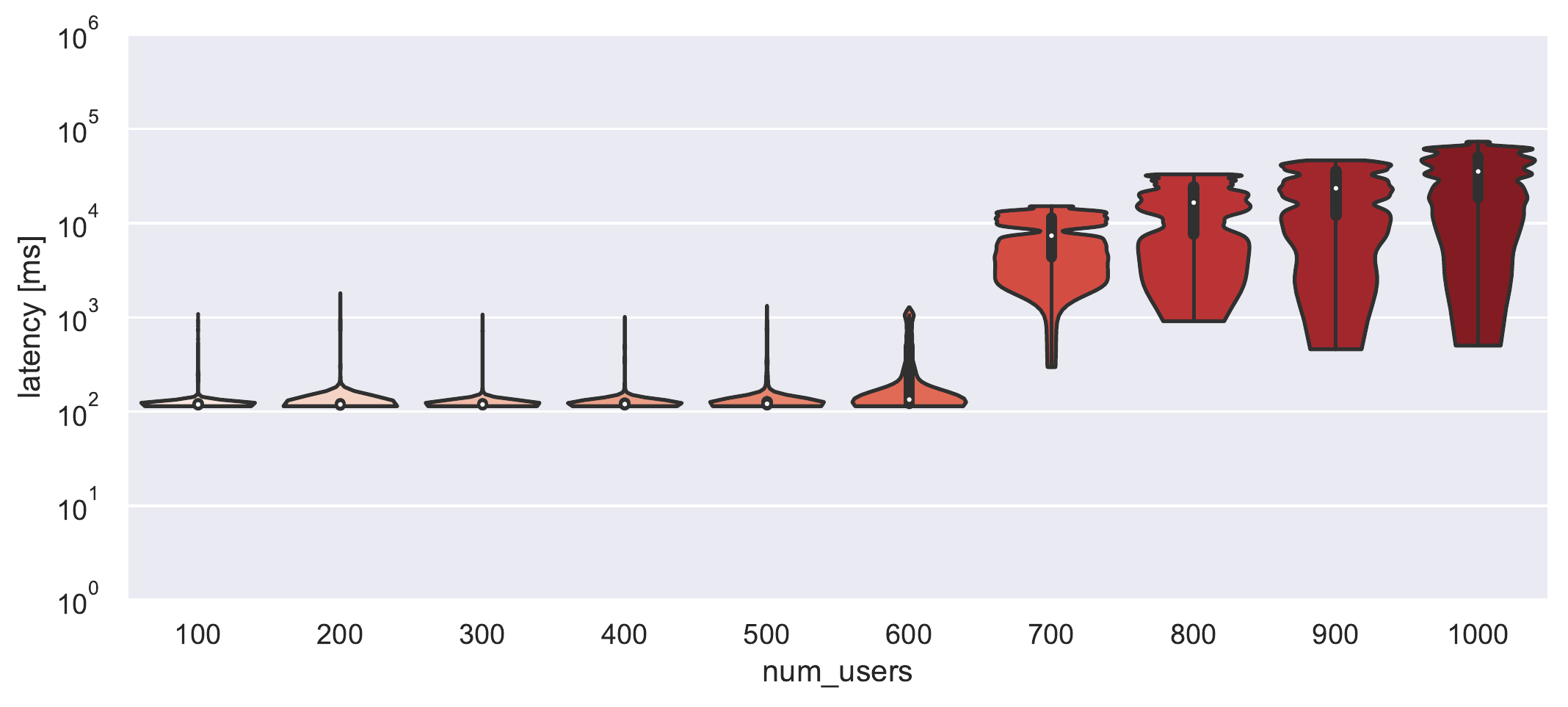}
	\label{fig_edge_cloud_violinplot_1KB_latency}}
	\caption{Latency with 1KB versus num. of users and scenarios.} 
	\vspace{-0.1cm}
	\label{latency_violin_1KB}
\end{figure}

\begin{figure}[ht]
	\centering
	\subfloat[cloud-only - 10KB payload size]{\includegraphics[width=0.99\columnwidth]{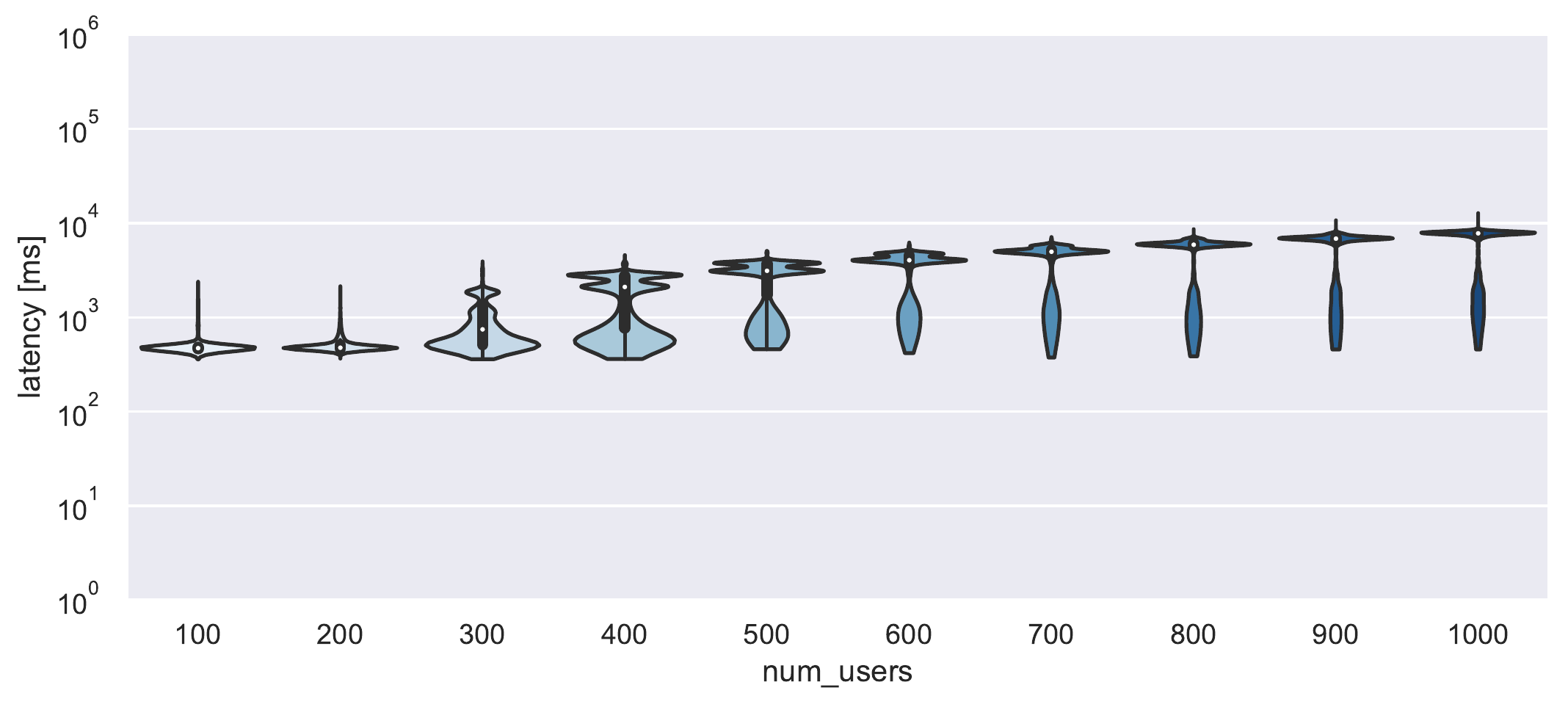}
	\label{fig_cloud_only_violinplot_10KB_latency}}
	\hfil
	\subfloat[edge-only - 10KB payload size]{\includegraphics[width=0.99\columnwidth]{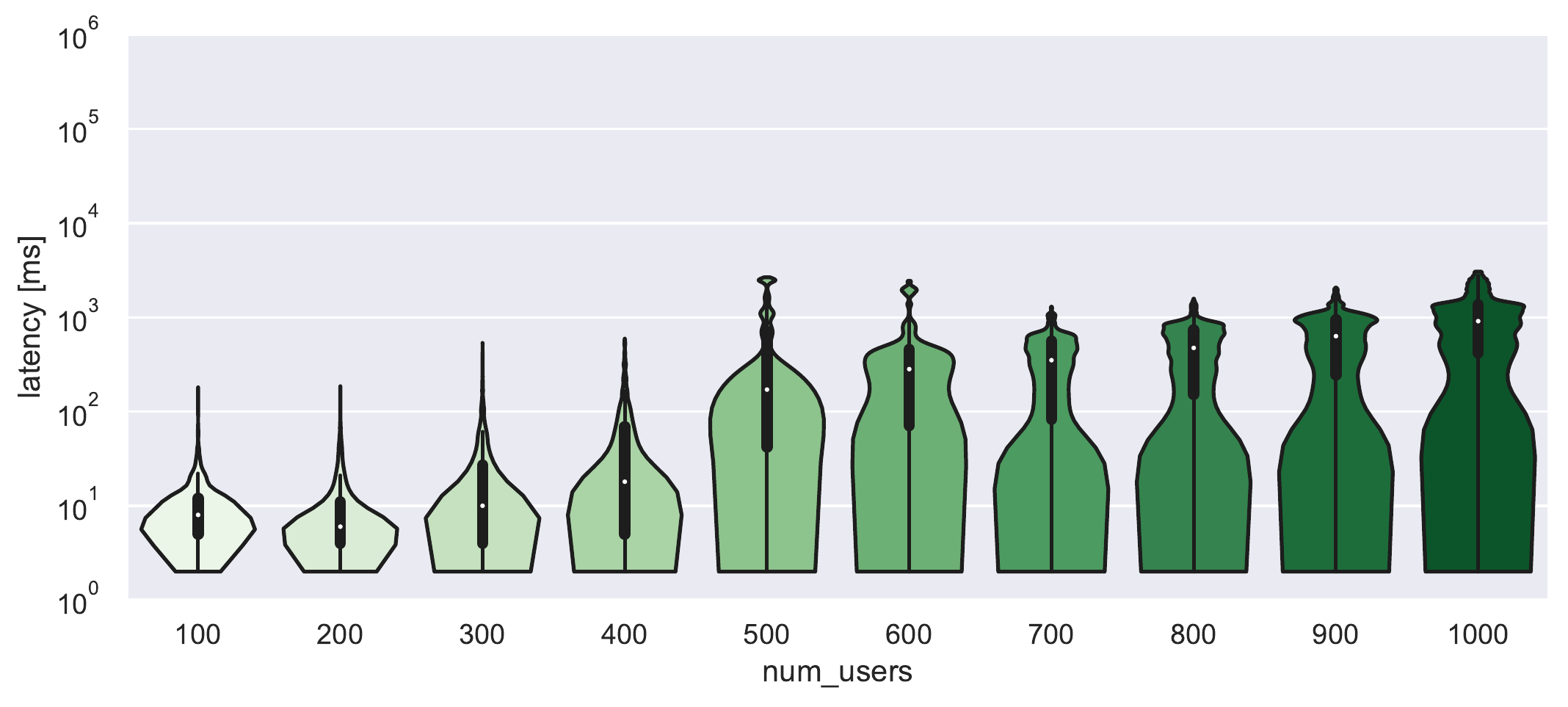} 
	\label{fig_edge_only_violinplot_10KB_latency}}
	\hfil
	\subfloat[edge-cloud - 10KB payload size]{\includegraphics[width=0.99\columnwidth]{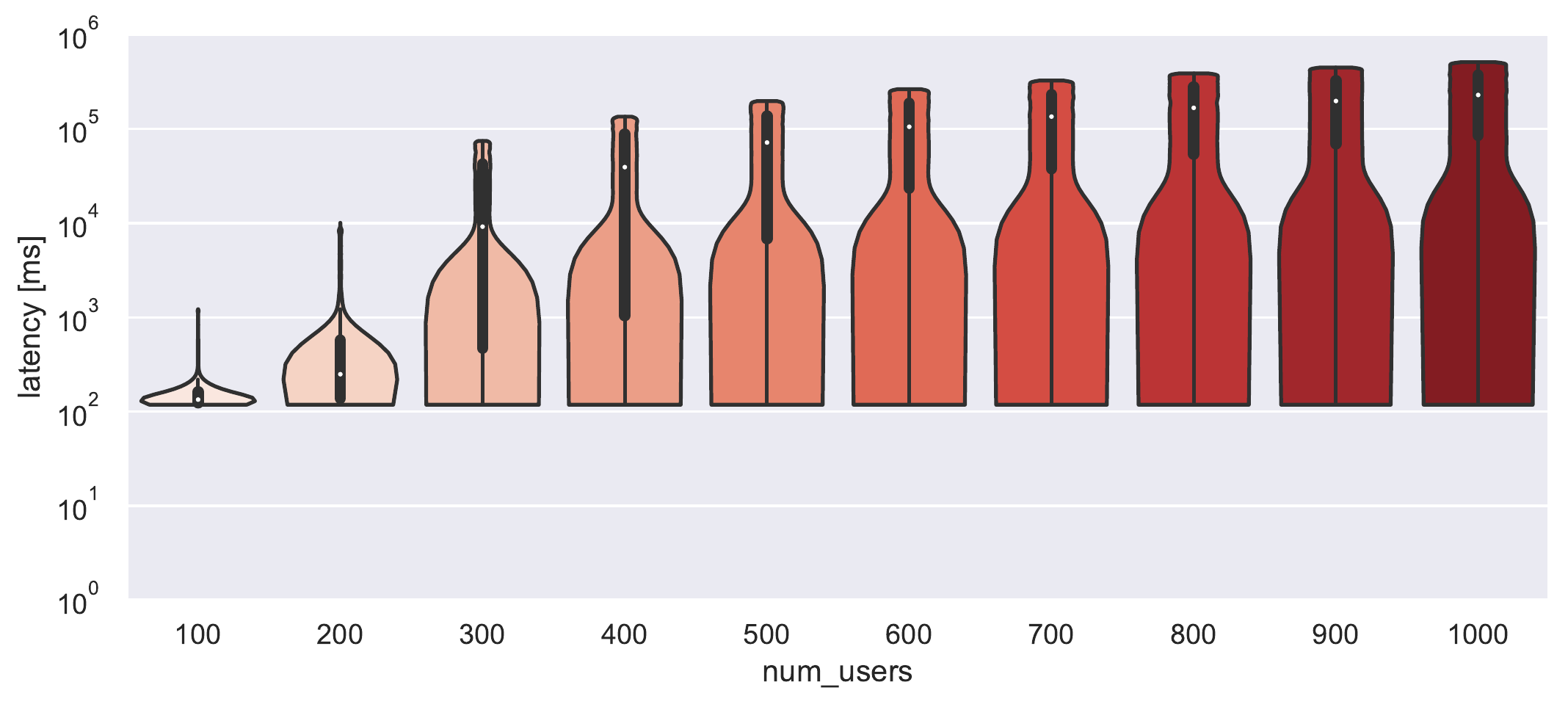}
	\label{fig_edge_cloud_violinplot_10KB_latency}}
	\caption{Latency with 10KB versus num. of users and scenarios.}
	\vspace{-0.3cm}
	\label{latency_violin_10KB}
\end{figure}

To show the latency results, we use violin plots (which represents the
probability density of the data as well as makers for median and interquartile
range when possible) and cumulative distribution function (CDF). We then compare
the results for all three scenarios previously described. Fig.
\ref{latency_violin_1KB} and Fig. \ref{latency_violin_10KB} show the latency of
the system for different number of users (i.e., processes) when performing
requests with 1 KB and 10 KB of payload data, respectively, in the scenarios
described in Fig. \ref{scenarios}.

\begin{figure}[ht]
	\centering
	\subfloat[1KB cloud-only]{\includegraphics[width=0.48\columnwidth]{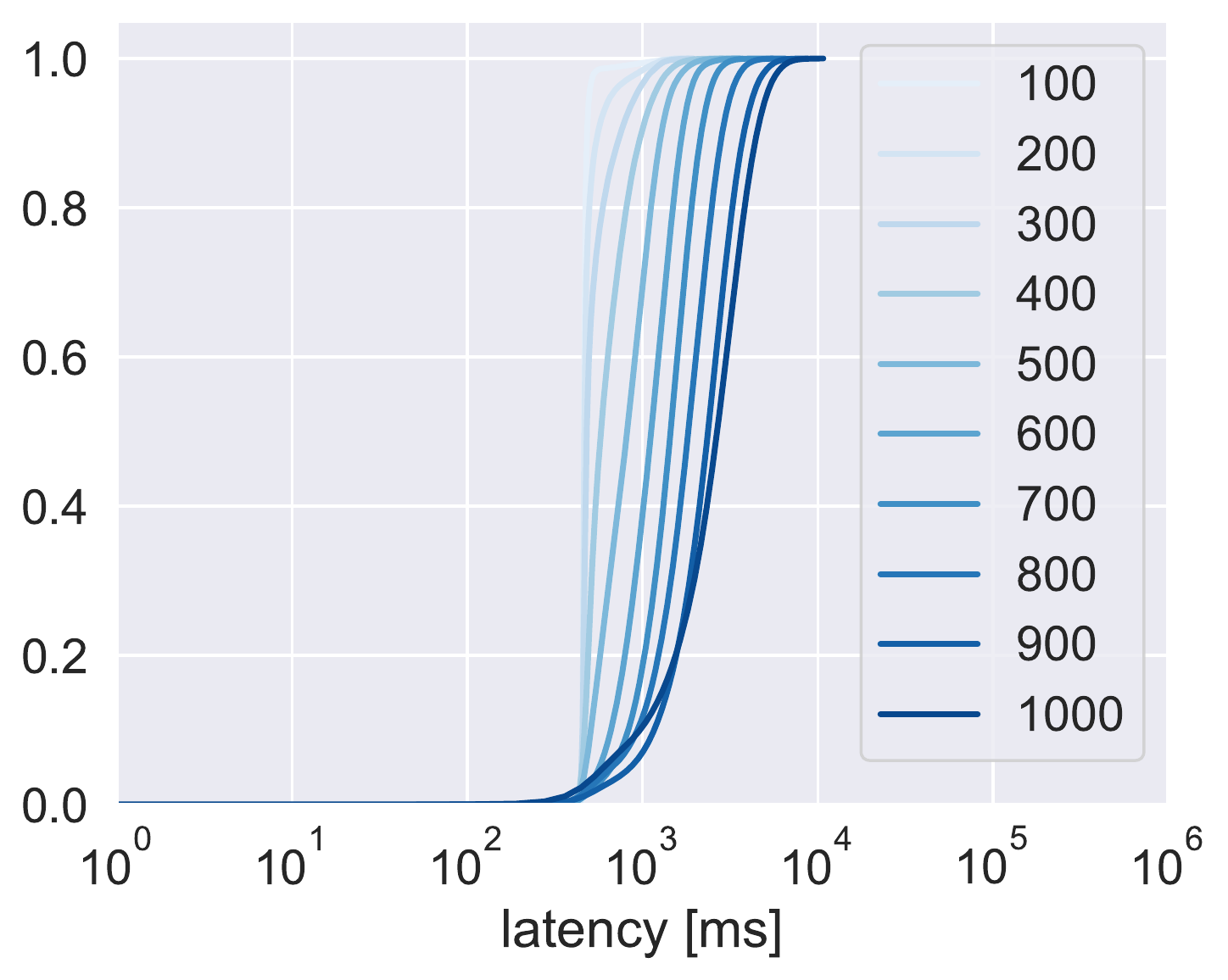} 
	\label{fig_cloud_only_cdf_1KB_latency}}
	\hfil
	\subfloat[10KB cloud-only]{\includegraphics[width=0.48\columnwidth]{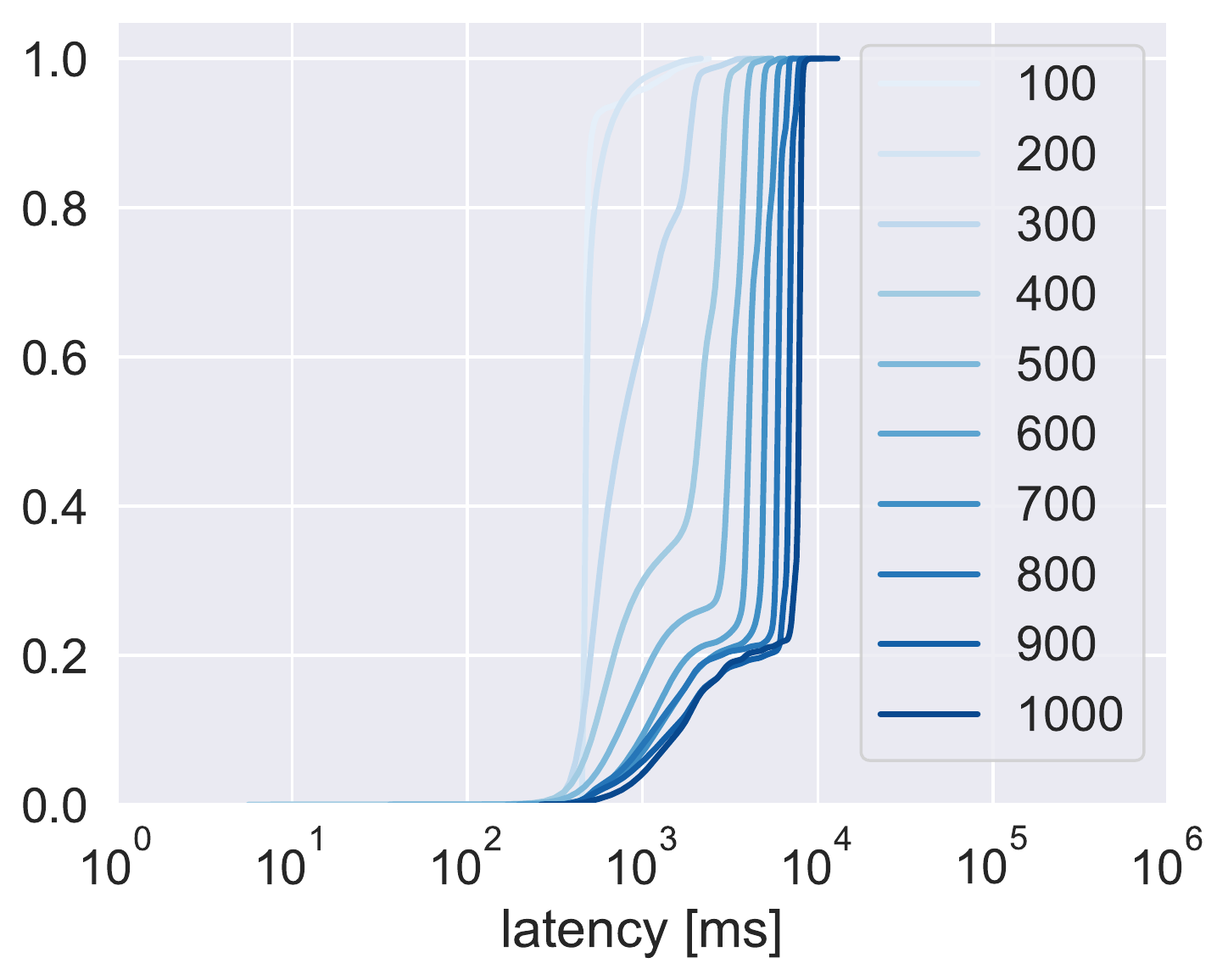}
	\label{fig_cloud_only_cdf_10KB_latency}}
	\hfil
	\subfloat[1KB edge-only]{\includegraphics[width=0.48\columnwidth]{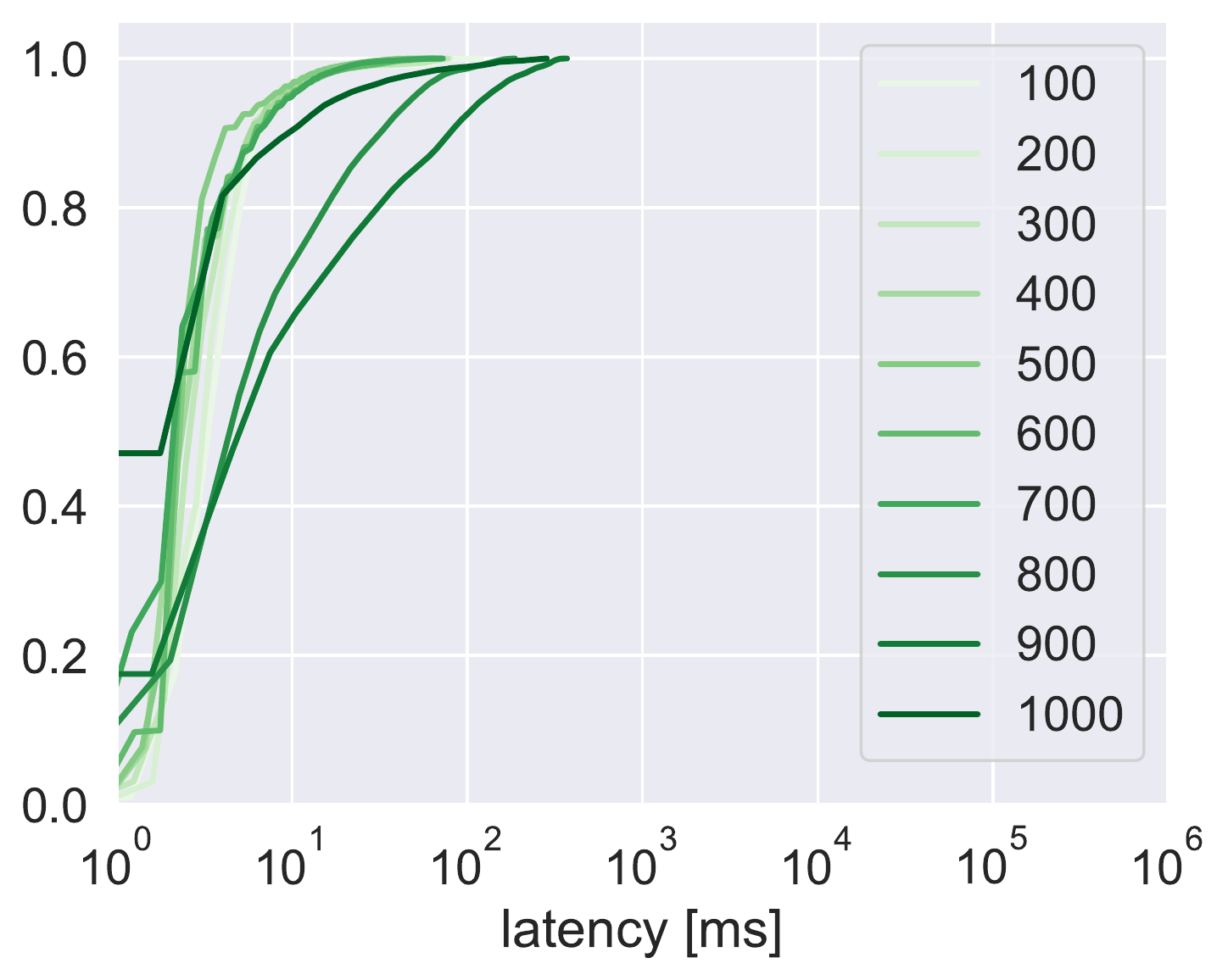} 
	\label{fig_edge_only_cdf_1KB_latency}}
	\hfil
	\subfloat[10KB edge-only]{\includegraphics[width=0.48\columnwidth]{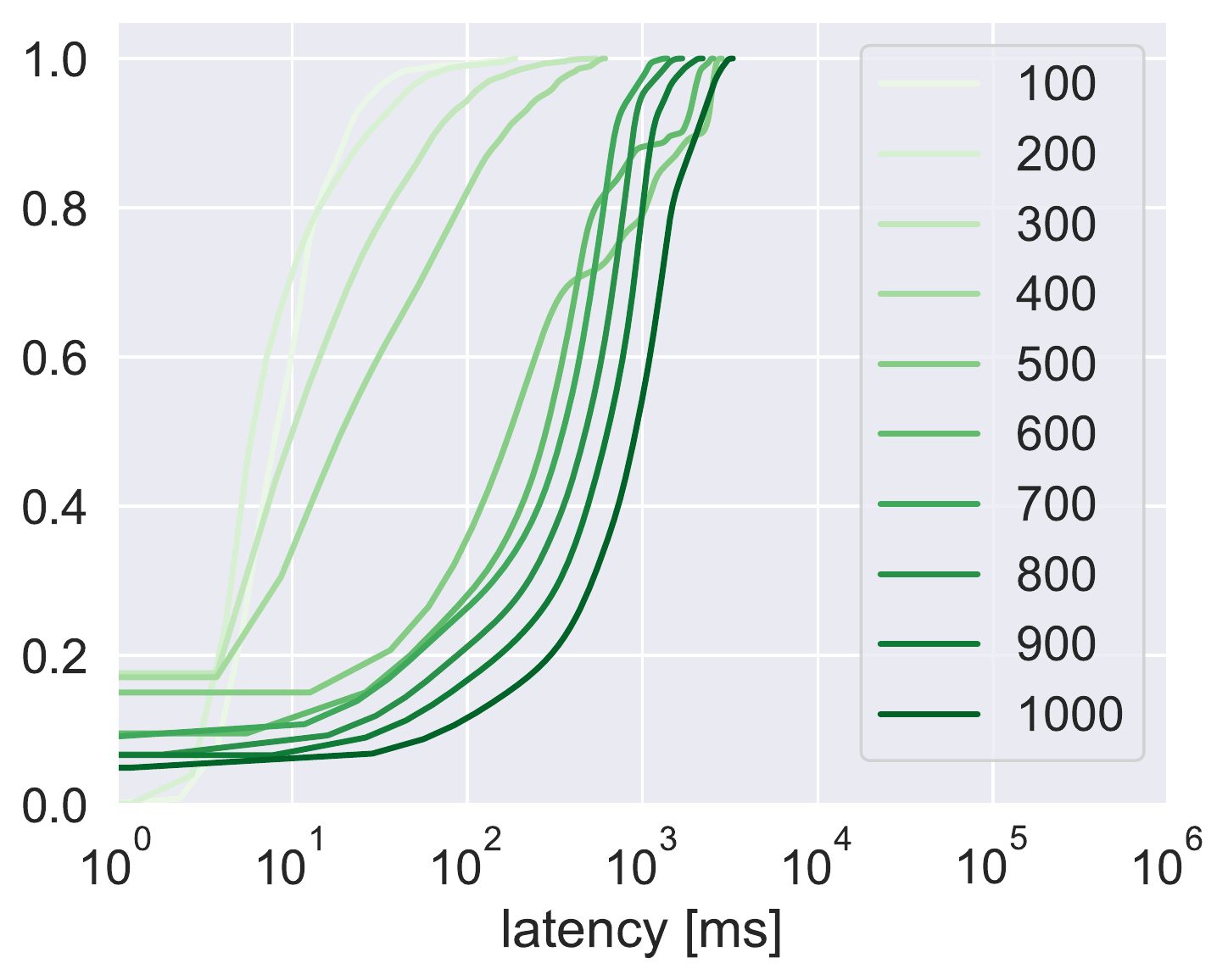}
	\label{fig_edge_only_cdf_10KB_latency}}
	\hfil
	\subfloat[1KB edge-cloud]{\includegraphics[width=0.48\columnwidth]{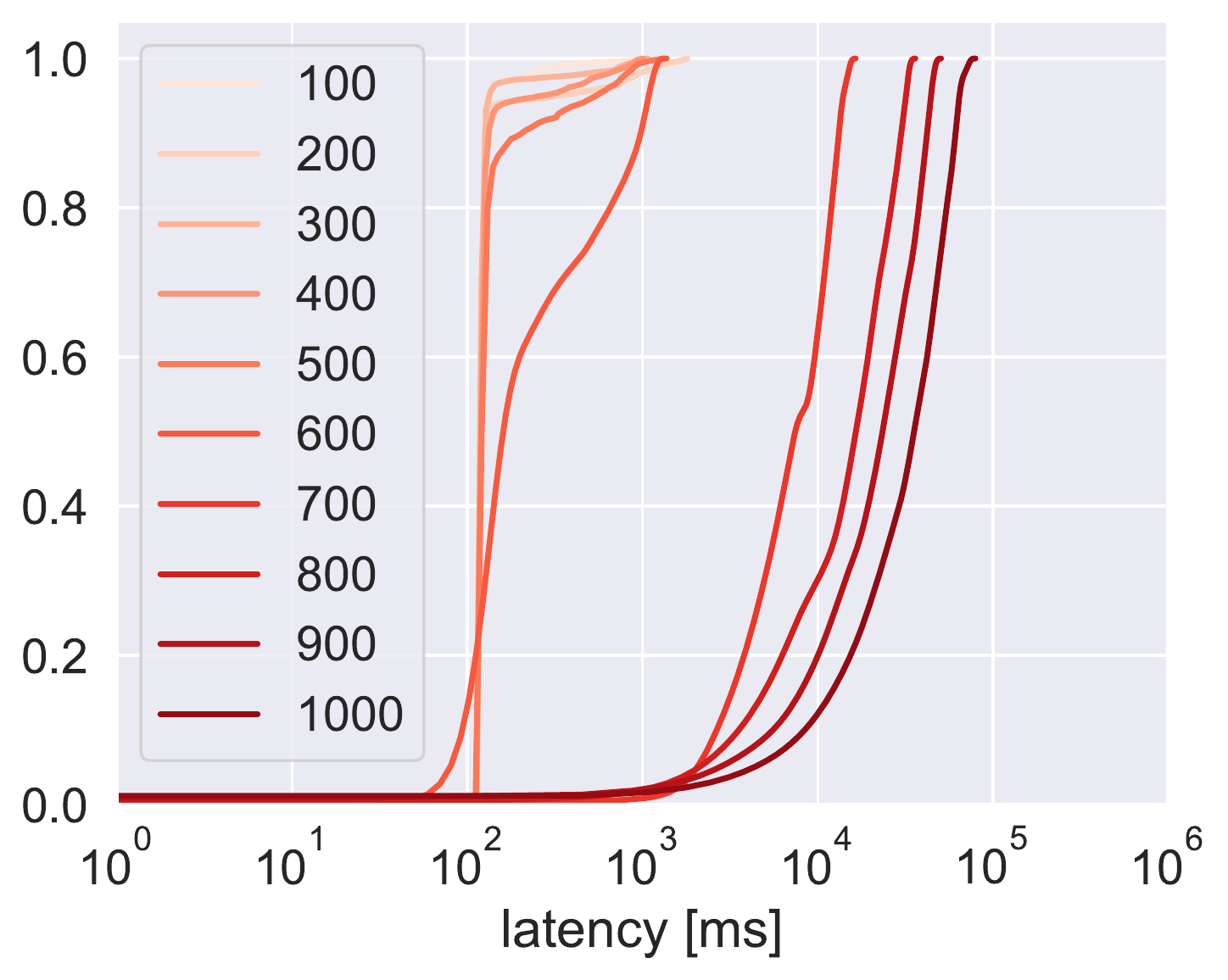} 
	\label{fig_edge_cloud_cdf_1KB_latency}}
	\hfil
	\subfloat[10KB edge-cloud]{\includegraphics[width=0.48\columnwidth]{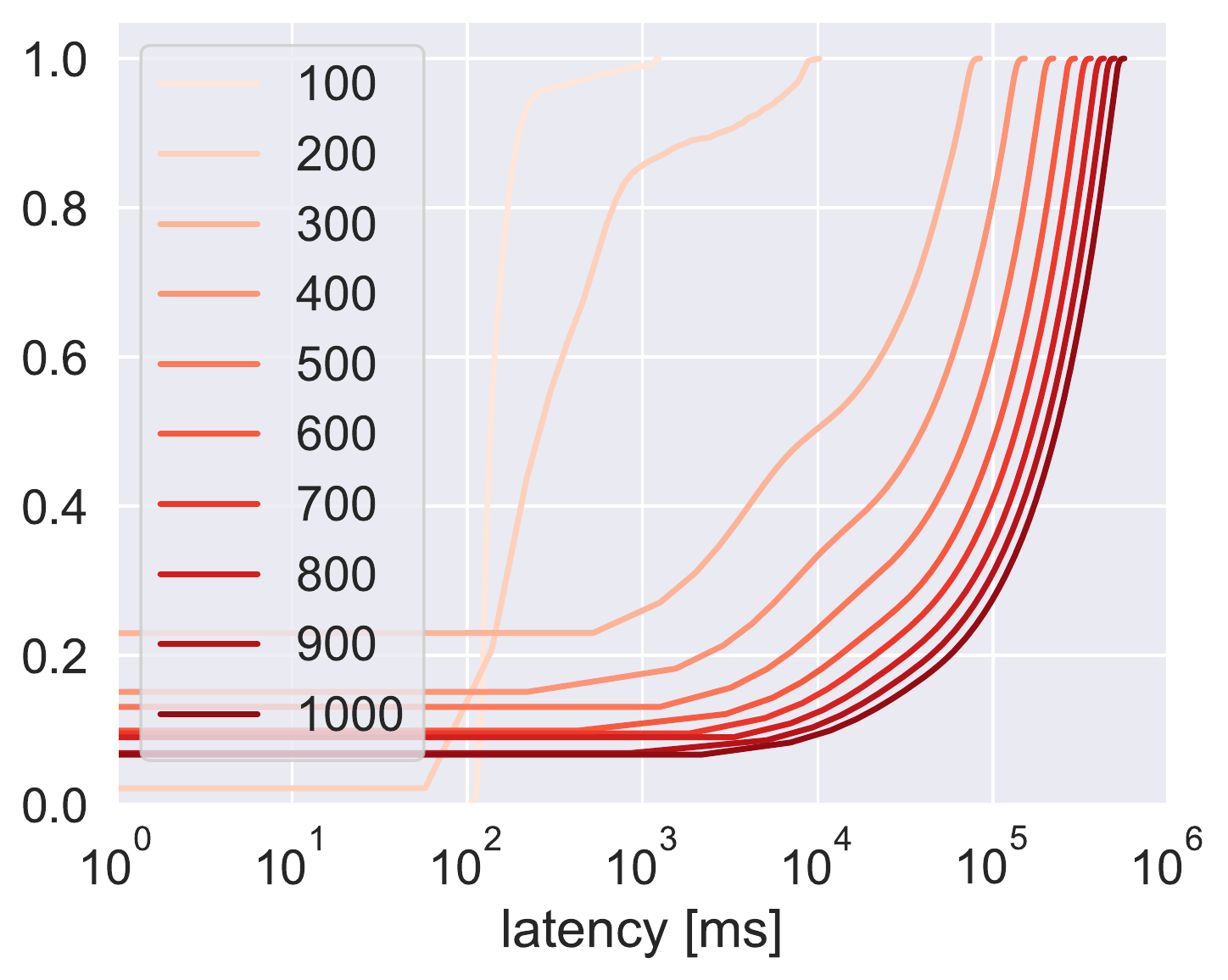}
	\label{fig_edge_cloud_cdf_10KB_latency}}
	\caption{CDF for latency for different number of users, payload size and scenarios.}
	\label{latency_cdf}
	\vspace{-0.2cm}
\end{figure}

Starting with the cloud-only scenario, Fig.
\ref{fig_cloud_only_violinplot_1KB_latency} and Fig.
\ref{fig_cloud_only_violinplot_10KB_latency} show the violin plots when the
payload data size is 1KB and 10KB, respectively. From these measurements, we can
observe that the increment in the number of users in the system increases
exponentially the average latency. This trend is more evident above 500
concurrent users when sending 1KB and 300 users when sending 10KB. The latency
distribution, with 400 users or less, when using 1KB, and with 200 users or
less, when using 10KB, is similar in both cases. The main point to note here is
the fact that above 500 and 300 users, respectively, the points are less
concentrated around the median, but spread between the maximum and minimum
values. This behavior can be explained by the fact that the cloud-only scenario
starts getting overloaded at the values of around 300 concurrent users, and the
response times may vary for every request, independently. Since this behavior is
more or less similar by either sending 1KB or 10KB, the reason behind is the way
that Firebase manages concurrent HTTP requests and less so because of the amount
of data sent. The difference between the amount of data sent can be, however,
appreciated by comparing the average values between both cases of data size,
being 10KB size clearly higher than with 1KB. The final point to consider from
these results is the fact that the results for 10KB are highly polarized as
compared to the case with 1KB. This shows how sending more data concurrently
with every request impacts on the variation of the response time of a request.
This behavior can be better observed by comparing Fig.
\ref{fig_cloud_only_cdf_1KB_latency} and Fig.
\ref{fig_cloud_only_cdf_10KB_latency} where the CDF shows how the latency for
10KB case is clearly affected above 300 users, which is not for the case of 1KB
of data. 

\par Fig. \ref{fig_edge_only_violinplot_1KB_latency} and Fig.
\ref{fig_edge_only_violinplot_10KB_latency} shows the latency results when
considering the edge-only scenario described in Fig. \ref{scenario2}. In this
case, the results for 1KB of payload data size do not exhibit large differences
with vairying number of users. For the case of data with 10 KB, there is clear
increment of the median above 400 users. The latter case also affects the
distribution of the latency measurements. Again, this behavior can be better
observed by comparing Fig. \ref{fig_edge_only_cdf_1KB_latency} and Fig.
\ref{fig_edge_only_cdf_10KB_latency} where for more than 400 users, the CDF is
more affected as compared to the case when only 1KB of data is sent. In Fig.
\ref{fig_edge_cloud_violinplot_1KB_latency} and Fig.
\ref{fig_edge_cloud_violinplot_10KB_latency}, we again show the latency, but in
this case for the combined edge-cloud scenario described in Fig.
\ref{scenario3}. Here, the behavior is quite different depending of the payload
size and for different number of users. For 1KB size, the results between 100
and 600 users are similar, and above 600, the latency exponentially increases
with the number of users (i.e., processes). For the case of sending 10KB, the
same behavior canbe observed but for over 200 users. The latter case is
interesting since the results are not concentrated around the median, but spread
out between the minimum and maximum values. Again, in Fig.
\ref{fig_edge_cloud_cdf_1KB_latency} and Fig.
\ref{fig_edge_cloud_cdf_10KB_latency} we can observe better the difference in
the the number of users affected in both cases.

\subsection{Scalability}

By comparing all three scenarios, we can observe that edge-only case scales
better than the other two cases for any number of users, with around 1 second of
latency in the worst case scenario. This is because, while in this case the data
is also synchronized to the cloud, the tests are being measured just between the
tester node and the edge node. Therefore here, we see a clear advantage of using
fast data stream tools such as Kafka, which can heavily reduce the  latency as
perceived by the end user, with data synchronized with the cloud. This, however,
does not show, for instance, the latency that two users located in different
locations would perceive when the data has to travel over the network. This case
can be only measured by using either cloud-only or edge-cloud scenarios where
the latency is measured after cloud synchronization. Then, by comparing these
two cases, we see how edge-cloud outperforms cloud-only when the system has 600
users or less for 1KB of data, and 200 or less when for 10KB of data. This
behavior is quite interesting since it would be more natural to expect that
edge-cloud is always slower than cloud-only by the sheer fact the data is
forwarded using Kafka and the Aggregator, which intuitively adds extra latency.
The reasoning behind this interesting behavior is that the Aggregator and Kafka
act as a buffer for the data and are more optimally synchronized with the cloud,
at least until certain amount of concurrent request and amount of data. Above
that level (700 users sending 1KB or 300 sending 10KB), edge-cloud does not
scale anymore, getting latency values between 10 and 100 seconds for 1KB, and
between 10 seconds and 1000 seconds for 10KB. In the cloud-only case, the system
scales well for 1KB case up to 600 users, and for 10KB case up to 300 users.
Above these values, the latency can reach up to 10 seconds, which is the point
where we can assume that the system does not scale anymore and becomes
practically unusable.

\subsection{Resource Utilization}

\begin{figure}[ht]
	\centering
	\subfloat[CPU utilization]{\includegraphics[width=0.48\columnwidth]{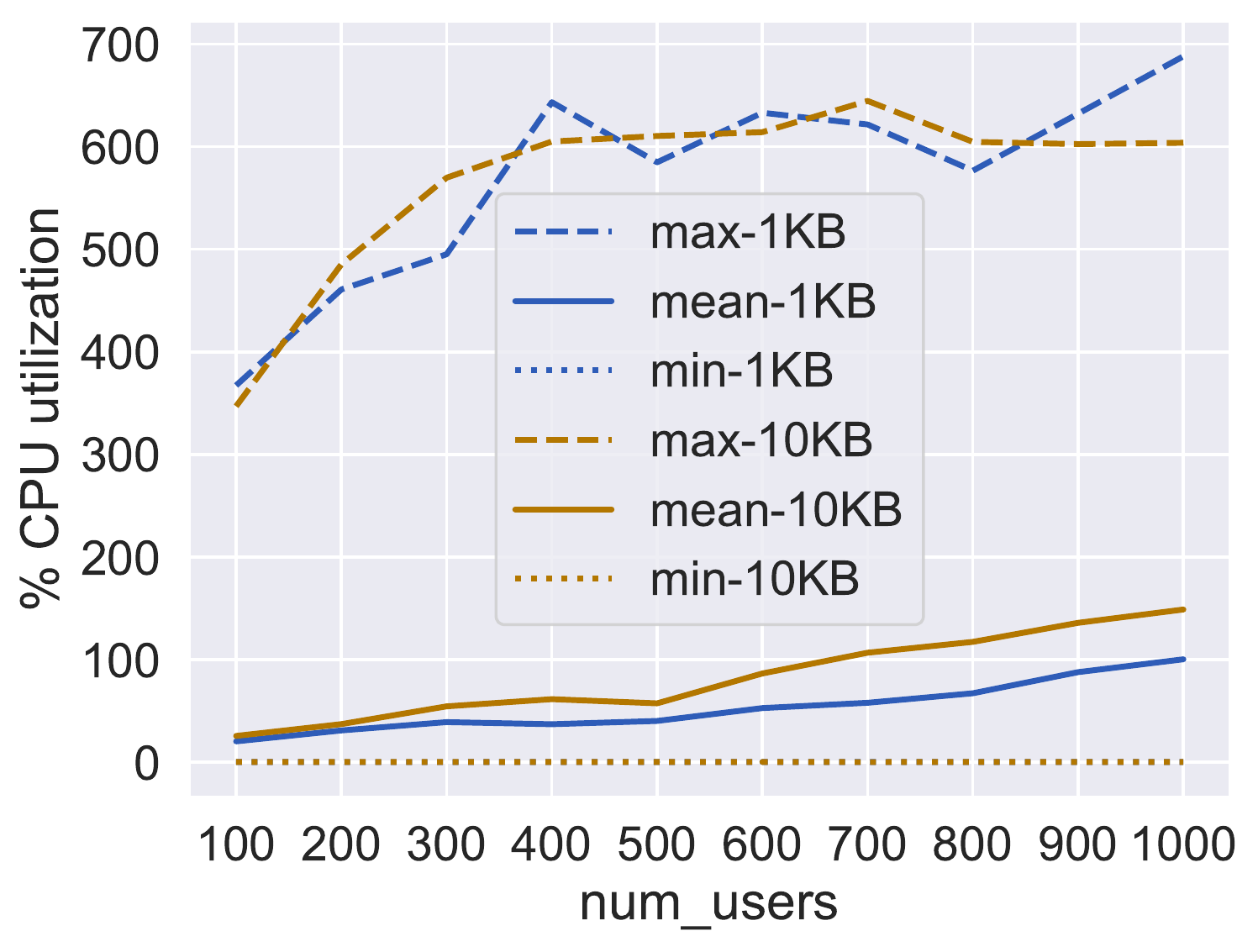}
	\label{fig_cpu}}
	\hfil
	\subfloat[Memory utilization]{\includegraphics[width=0.48\columnwidth]{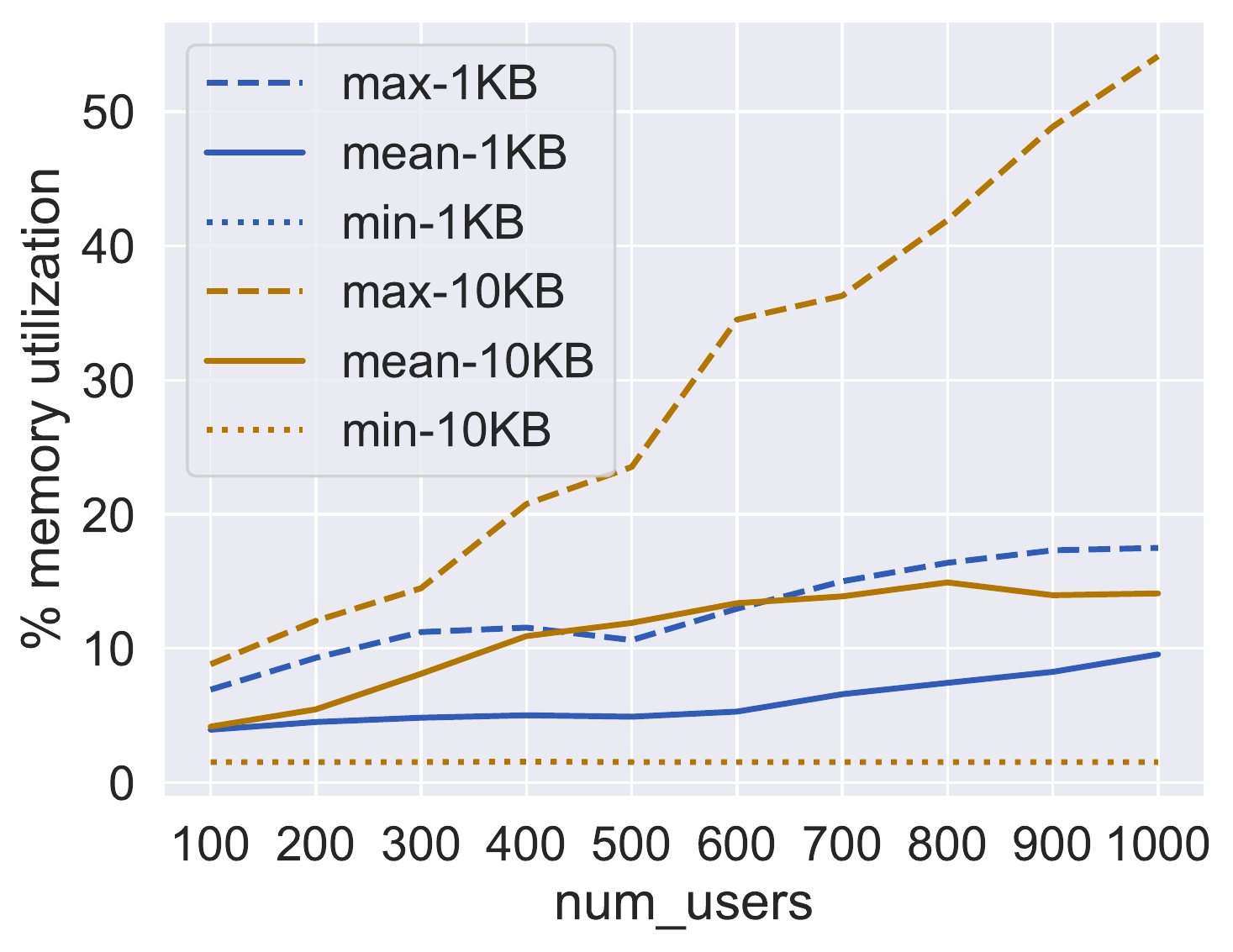}
	\label{fig_memory}}
	\hfil
	\subfloat[Network I/O]{\includegraphics[width=0.48\columnwidth]{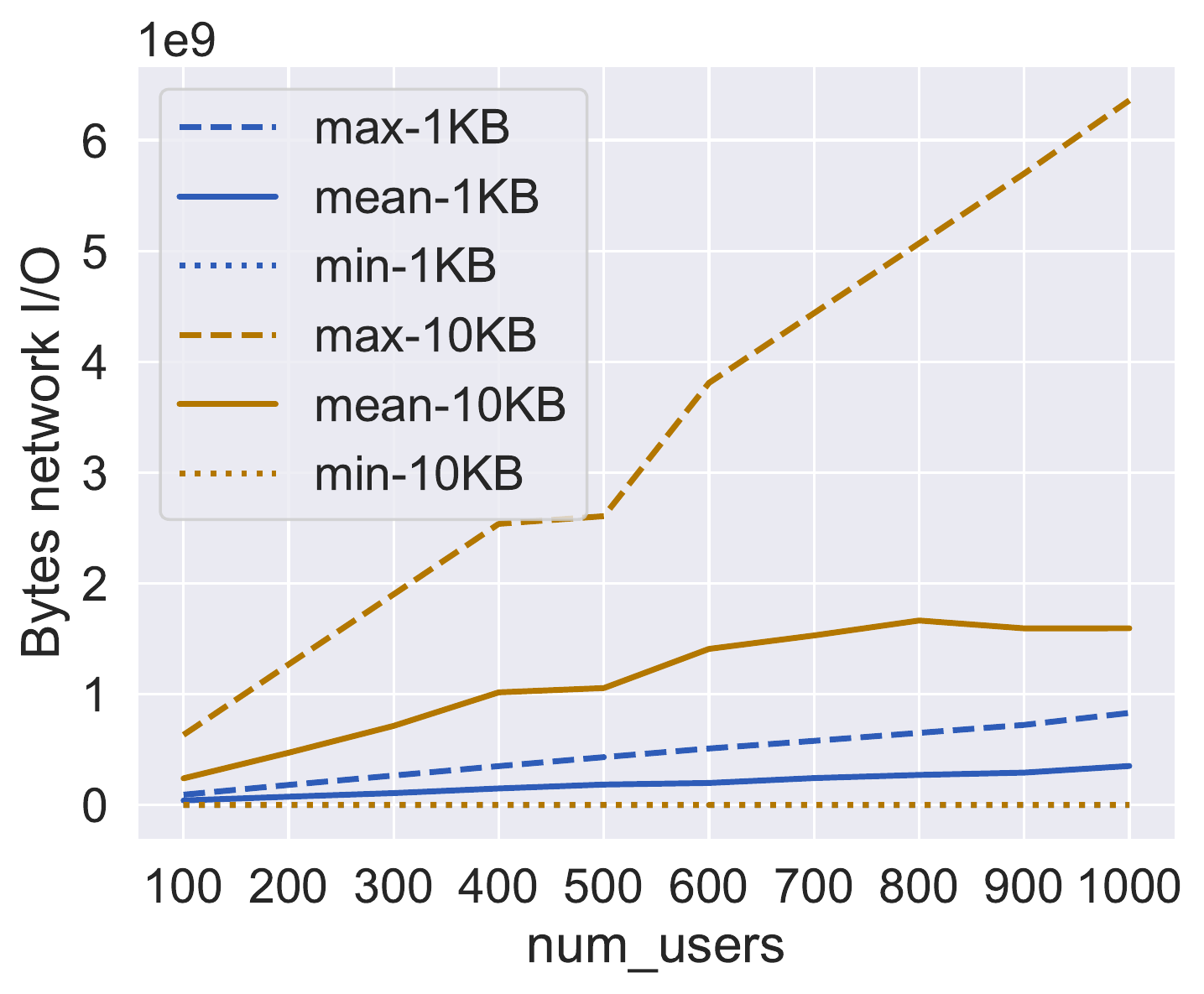}
	\label{fig_netI}}
	\caption{Resource utilization at the edge node.}
	\label{resource_utilization} 
	\vspace{-0.2cm}
\end{figure}

Since the objective of our experimental study is also to show the feasibility of
edge-cloud systems realized with common-purpose hardware, we now provide
measurements on resources utilized by the edge node when running the tests.
Because the edge node is running three different containers, we use Docker stats
command to retrieve the the status of all three containers which are combined
and stored in an output file every second. Fig. \ref{fig_cpu} shows the mean,
maximum and minimum CPU utilization for all number of users and both payload
data sizes. It should be noted here that Docker considers 100\% utilization when
1 core is fully utilized, so results above 100\% mean that more cores have been
used. Here, we can see how the mean value slightly increases with the number of
users, with the case of 10KB being larger than 1KB in all cases. The maximum
values are in both cases similar, and are progressively increasing from 100 to
400 users, and later stabilized at around 600 users. The memory utilization is
shown in Fig. \ref{fig_memory}. Here we can observe that the mean value slightly
increases with the number of users, but the maximum value for the case of 10KB
increases heavily with the number of users. This is due to the excessive memory
usage that the Aggregator requires from the moment of the broker receiving the
data until the data is stored into the database and removed from the memory.
Since these intervals are short in time, they affects the maximum values but not
to the mean values. Fig. \ref{fig_netI} shows the results for the network
input/output (I/O), whereby in this case both input and output are the same.
This is due to all three containers combined, and for all data received are sent
out again. In this case, we can observe the pattern similar to the memory
measurements, whereby the maximum value for the 10KB case heavily increases with
the number of users, from around 1GB of data transmitted to 6GB, which is in
line to what we could also observe in memory measurements.

\section{Conclusions}

While edge computing is expected to be the solution for latency sensitive
applications that require high intensive processing tasks, it is always an open
question of whether a real-world edge computing system implementation can
achieve the performance forseen. We engineered a modular edge cloud computing
system architecture using the latest advances on containerization platforms and
open source tools and show the challenges and importance of benchmarking
latency, scalability and resource utilization in edge computing. We compared
experimentally three scenarios: cloud-only, edge-only and edge-cloud. The
measurements showed that while edge-only outperforms other cases in terms of
latency and scalability, it also requires the app to work with data located at
centralized edge nodes. Edge-cloud performs around 10 times better compared to
only-cloud until certain number of concurrent processes where the system does
not scale anymore, and only-cloud performs better. Finally, for resource
utilization the maximum memory and network I/O increase heavily with increasing
amounts of data and concurrent users. 

\bibliographystyle{IEEEtran}
\bibliography{icc-2020}

\end{document}